# miRNA and Gene Expression based Cancer Classification using Self-Learning and Co-Training Approaches

Rania Ibrahim[1], Noha A. Yousri[2,1], Mohamed A. Ismail[1] and Nagwa M. El-Makky[1]
rania.ibrahim.salama@gmail.com, noha.yousri@alexu.edu.eg, drmaismail@gmail.com and
nagwamakky@alexu.edu.eg
[1]Computer and Systems Engineering Department, Alexandria University.
Alexanrdria 21544, Egypt.
[2]Computer Science and Engineering, Egypt-Japan University of Science & Technology (E-JUST).

*Abstract*— miRNA and gene expression profiles have been proved useful for classifying cancer samples. Efficient classifiers have been recently sought and developed. A number of attempts to classify cancer samples using miRNA/gene expression profiles are known in literature. However, the use of semi-supervised learning models have been used recently in bioinformatics, to exploit the huge corpuses of publicly available sets. Using both labeled and unlabeled sets to train sample classifiers, have not been previously considered when gene and miRNA expression sets are used. Moreover, there is a motivation to integrate both miRNA and gene expression for a semi-supervised cancer classification as that provides more information on the characteristics of cancer samples.

In this paper, two semi-supervised machine learning approaches, namely self-learning and co-training, are adapted to enhance the quality of cancer sample classification. These approaches exploit the huge public corpuses to enrich the training data. In self-learning, miRNA and gene based classifiers are enhanced independently. While in co-training, both miRNA and gene expression profiles are used simultaneously to provide different views of cancer samples. To our knowledge, it is the first attempt to apply these learning approaches to cancer classification. The approaches were evaluated using breast cancer, hepatocellular carcinoma (HCC) and lung cancer expression sets. Results show up to 20% improvement in F1-measure over Random Forests and SVM classifiers. Co-Training also outperforms Low Density Separation (LDS) approach by around 25% improvement in F1-measure in breast cancer.

## I. INTRODUCTION

MicroRNAs (miRNAs) are short (19–25 nucleotides) noncoding single-stranded RNA molecules [1], which are cleaved from 70–100 nucleotide miRNA precursors. miRNAs regulate gene expression either at the transcriptional or translational level, based on specific binding to the complementary sequence in the coding or noncoding region of mRNA transcripts [1]. Recent research has pointed out the success of using miRNA and gene expression datasets in cancer classification; miRNA profiles were used recently to discriminate malignancies of the breast [2], lung ([2], [3]), pancreas ([2], [4]) and liver ([5], [6], [7]). Enhancing the accuracy of cancer classifiers, based on miRNA, gene, or mRNA expressions, has been targeted in previous work ([8], [9], [10], [11], [12], [13]). Moreover, different feature selections and classification methods that efficiently detect the malignancy status (normal or cancer) of the tissues were previously explored in [11]. In addition, two classifiers are built [12], one for miRNA data and another for mRNA data. The main drawbacks of the approach is that it assumes paired miRNA and mRNA data for each patient and it uses decision fusion rule to combine the classifiers decision without enhancing the classifiers themselves. Random Forests have been used in classifying cancer in [14], [15] and [16]. Also, SVM has been used in classifying cancer as in [17] and [18].

The idea of combining both labeled and unlabeled sets using semi-supervised machine learning approaches has been used to enhance classifiers in other domains like object detection [22], word sense disambiguation [23] and subjective noun identification [24]. Semi-supervised learning also has proved to be effective in solving several biology problems like protein classification [25] and prediction of factor-gene interaction [26]. However, in the field of sample classification using gene and miRNA expression, semi-supervised machine learning techniques were not considered before. Microarray experiments are time consuming, expensive and limited, that is why usually the number of samples of microarray-based studies is small [27]. Thus, huge publicly available gene/miRNA expression sets with unlabeled samples are tempting to use for enriching training data of sample classifiers. Integrating both miRNA and mRNA expression profiles were thought to provide complementary information [12], as miRNAs regulate gene expression at the post-transcriptional level. In co-training, both miRNA and gene expression profiles are used simultaneously to provide different views of cancer samples. Semi-supervised machine learning approaches are applied in this paper to discriminate cancer subtypes. Discriminating cancer subtypes helps in understanding the evolution of cancer and is used to find appropriate therapies. For example, angiogenesis inhibitors like bevacizumab are more effective in treating adenocarcinoma lung cancer than squamous phenotypes ([19], [20]). Also, breast cancer has an unpredictable response, and developing effective therapies remain a major challenge in the clinical management of breast cancer patients [21]. Moreover, identifying metastasis hepatocellular carcinoma (HCC) samples is an important task as metastasis is a complex process that involves multiple alterations ([39], [40]).

In this paper, two semi-supervised machine learning approaches, namely self-learning [28] and co-training ([29], [30]) are used to enhance the classification accuracy of cancer samples by combining both labeled and unlabeled miRNA and gene expression profiles. In self-learning, a classifier is initially constructed using the labeled set, then its accuracy is enhanced by adding more data from unlabeled sets. Self-learning is used on one particular source of expression, i.e either gene or miRNA expression data. In co-training, two classifiers are trained, each is specific to a different source of expression data (gene or miRNA), termed as two views of the data. Based on the two views, two classifiers are constructed and then used to train each other. Exchanging information between the two classifiers requires a mapping from miRNA expression to gene expression or the opposite. A simple mapping is thus suggested based on known biological relations between miRNAs and their target genes.

The aforementioned classification approaches were evaluated using gene and miRNA expression profiles of three different cancer types: breast cancer, hepatocellular carcinoma (HCC) and lung cancer. The results show around 20% improvement in F1-measure in breast cancer, around 10% improvement in precision in metastatic HCC cancer and 3% improvement in F1-measure in squamous lung cancer over the Random Forests and SVM classifiers. Also, the approaches were compared to another semi-supervised approach called Low Density Separation (LDS), which was used to enhance the classifiers of cancer recurrence in [27]. The results show that co-training outperforms LDS by exploiting the two different views, i.e. miRNA expression view and gene expression view.

The paper is organized as follows section II discusses the related work, while section III describes the proposed approaches in details and section IV shows experimental results. Finally section V concludes the paper.

## II. RELATED WORK

Using miRNA expression profiles to discriminate cancerous samples from normal ones, and to classify cancer into its subtypes, is an active research area and was applied to different cancer types as breast [2], lung ([2], [3]), pancreatic in ([2], [4]) and liver in ([5], [6], [7]). The previous papers used one of the following supervised machine learning techniques like SVM, Prediction Analysis of Microarrays (PAM) and compound covariate predictor.

Several attempts for enhancing cancer classifiers have been recently introduced ([11], [12], [13]). In [11], number of feature selection methods, as Pearson's and Spearman's correlations, Euclidean distance, cosine coefficient, information gain and mutual information and signal-to-noise ratio are used to enhance cancer classifiers. Also different classification methods which are k-nearest neighbor methods, multilayer perceptrons, and support vector machines with linear kernel are used [11]. The work has focused only on improving classifiers based on labeled samples miRNA expression profiles and didn't use publicity available unlabeled sets, also, gene expression profiles were not used to enhance miRNA based cancer samples classifiers. Enhancing the classification accuracy by building two classifiers one for miRNA data and another for mRNA data were explored in [12]. That work first applies feature selection using relief-F feature selection, then it uses bagged fuzzy KNN classifier and finally it combines the two classifiers using fusion decision rule. The drawback of the approach is that it assumes the existence of both miRNA and mRNA data for each patient and it just uses decision fusion rule to combine the classifiers decision without enhancing the classifiers themselves. Another work [13] has considered using discrete function learning (DFL) method on the miRNA expression profiles to find the subset of miRNAs that shows strong distinction of expression levels in normal and tumor tissues and then uses these miRNAs to build a classifier. The paper didn't combine multiple miRNA dataset or use gene expression dataset to enhance the classifier. Semi-supervised machine learning approaches were introduced in classification using expression sets by using LDS approach which was used in [27] to enhance cancer recurrence classifiers. Semi-supervised machine learning approaches make use of the publicity available unlabeled sets to enrich the training data of the classifiers. However, the approach depends only on gene expression, and didn't combine both miRNA and gene expression sets.

Other semi-supervised machine learning approaches like self-learning and co-training were introduced in other domains. The heuristic approach of self-learning (also known as self-training) is one of the oldest approaches in semi-supervised learning and that was introduced in [28]. Self-learning was used in many applications as object detection [22], word sense disambiguation [23] and subjective noun identification [24]. Also, co-training is a semi-supervised approach that appeared in [29] and [30] and is also used in applications as word sense disambiguation [31] and email classification [32].

In this paper, self-learning and co-training approaches are used. Both approaches use unlabeled sets to enhance classifiers accuracy. Co-training also enhances the results by combining both miRNA and gene expression sets. The results show improvements over Random Forests and SVM classifiers and LDS approach.

## III. SELF-LEARNING AND CO-TRAINING ADAPTATION TO MIRNA/GENE BASED CLASSIFICATION

In self-learning and co-training, the objective is to construct a classifier to discriminate between different cancer subtypes, given the following:

- The expression vector of a sample i, denoted $x_i$, is defined as follows:

$$x_i = \{e_{i1}, e_{i2}, \ldots, e_{ij}, \ldots, e_{iM}\}$$

    Where $e_{ij}$ is the expression value of the j$^{th}$ miRNA/gene, and M is the number of miRNAs/genes.

- N is the number of samples.

Two sets are used in both self-learning and co-training, which are defined as follows:

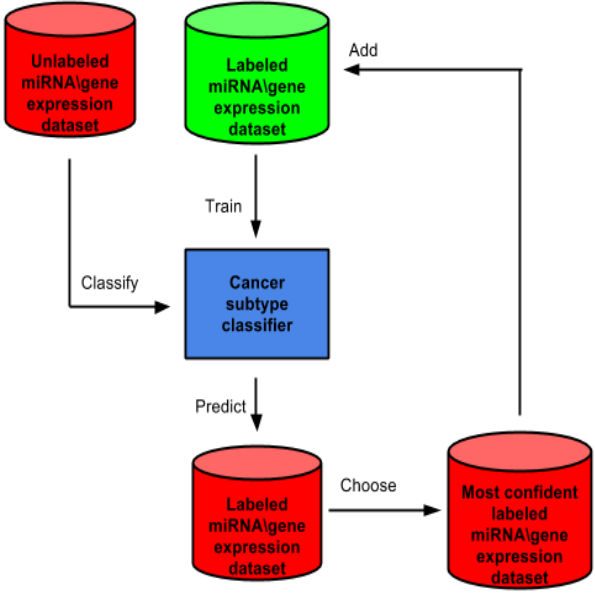

Fig 1. Self-Learning approach overview.

- A set of labeled samples L; $L = \{x_i, y_i\}_{i=1}^{N}$ Where $y_i$ is the cancer subtype label.
- A set of unlabeled samples U; $U = \{x_i\}_{i=1}^{N}$

The size of U is expected to be much larger than L (|U| >> |L|), which is expected to help enhancing the accuracy of the classifiers by adding more expression vectors to the training data. Increasing the number of unlabeled sets leads to higher enrichment in the training set. Moreover, increasing the overlap between the miRNAs/genes in the labeled and unlabeled sets leads to increasing the effect of adding the unlabeled sets.

Self-learning [28] is a semi-supervised machine learning approach, in which the labeled set L is used to build the initial classifier and the unlabeled set U is utilized to enhance its accuracy by adding the unlabeled samples with the highest classification confidence to the training set, thus resulting in making the classifier learns based on its own decision. Co-training ([29], [30]) is also a semi-supervised machine learning approach, which requires two views of the data. Two classifiers are constructed separately for each view. Each classifier is then used to train the other one by classifying unlabeled samples and train the other view with the samples with highest classification confidence.

The next sections explain how the two approaches are adapted to use the unlabeled set U to enhance the baseline classifier constructed based on L.

*A. Self-Leanring Adaptation*

The steps of adapting the self-learning approach are described as follows:

a) Train an initial cancer subtype classifier using set L.

b) Use the initial classifier to identify the subtype labels of the unlabeled set U.

c) Choose the most confident subset of cancer samples (U'), i.e. samples classified with a confidence greater than a given threshold (α).

d) Append the set of most confident samples to the initial training dataset to form a new training set (U' ∪ L) for re-training the classifier.

e) Use the classifier constructed at step d to perform several iterations over the unlabeled set(s). At each iteration, re-apply steps b, c and d.

The resulting classifier can then be used to classify new samples based on their miRNA/gene expression profiles. The confidence threshold α should be appropriately selected. Decreasing α can increase the false positives rate. On the other hand, increasing α can result in restricting the learning process to the highly confident samples, typically the ones that are most similar to the training data, thus losing the benefit of including more training samples to the labeled data. Tuning parameter α is thus important, since it affects the classifier's accuracy to choose the samples that will enhance the classifier.

The next section explains the co-training idea and adaptation in details.

*B. Co-Training Adaptation*

In this paper, the co-training approach is adapted to classify cancer subtypes by training two different classifiers; the first is based on the gene expression view and the second is based on the miRNA expression view. Each view captures a different perspective of the underlying biology of cancer and integrating them using the co-training pipeline exploits this information diversity to enhance the classification accuracy. The following steps describes co-training in details:

a) Two initial cancer classifiers are separately constructed; one from the miRNA expression dataset ($L_{miRNA}$) and another one from the gene expression dataset ($L_{gene}$) using manually labeled cancer subtypes sets.

b) Let the initial classifiers separately classify the unlabeled cancer miRNA/gene expression datasets ($U_{miRNA}$/ $U_{gene}$) into cancer subtypes.

c) Choose the most confident labeled subtypes samples ($U'_{miRNA}$ & $U'_{gene}$) that have classification scores greater than α.

d) Retrieve miRNA-gene relations using miRanda. For the classifiers to train each other, miRNA expression should be mapped to gene expression and vice versa. miRNAs and their target genes databases are used to map the datasets. In our case, miRanda [33] database is used.

e) Append the mapped miRNA expression sets to the gene expression training sets and the mapped gene expression sets to the miRNA expression training sets and re-train the classifiers.

f) Use the classifier constructed at step e to perform several iterations over the unlabeled set(s). At each iteration, re-apply steps b, c, d and e.

g) In step d, a mapping between the miRNA view and gene view is required. As shown in figure 2, miRNAs and their target genes are related by a many to many relationship; multiple miRNAs target the same gene, and multiple genes are targeted by the same miRNA. For the classifier to exploit the two views, i.e. gene and miRNA sets, a miRNA expression vector is constructed from its target gene's expression vector.. Due to the many to many relationship between miRNAs and genes, it is suggested to use an aggregation of all expression vectors of the target genes to represent the miRNA expression vector. Similarly, a gene expression vector is constructed by aggregating the expression vectors of the miRNAs that target this gene. To map a gene to a miRNA, or the opposite, it is proposed to take the mean expression value of all miRNAs related to a gene, or the opposite, i.e. the mean expression value of all genes related to a miRNA. Experimental results show that taking the mean value of expressions has improved the classification accuracy. Part of the future work would be investigating the effect of using other methods as a mapping function.

After the co-training process, the two classifiers can be used independently, one on gene expression profile and the other on miRNA expression profile of cancer samples. The next section shows the experimental results of both self-learning and co-training approaches.

## IV. EXPERIMENTAL RESULTS

Two core classifiers of self-learning and co-training were used, which are Random Forests and SVM. RF is a known classifier ensemble method [34] based on constructing multiple decision trees. Each decision tree is built on a bootstrap sample of the training data using a randomly selected subset of features. For predicting a sample's label, a majority vote based on the classification obtained from the different decision trees is calculated. RF have been used in classifying cancer in [14], [15] and [16]. RF implementation from the Weka repository [35] was used, and the number of decision trees was set to 10. SVM implementation was also used from the Weka repository [35].

The approaches were evaluated using three cancer types, namely breast cancer, hepatocellular carcinoma (HCC) and lung cancer. miRNA based classifiers were constructed for breast cancer and HCC sets, while gene based classifiers were constructed for all 3 sets. In addition, self-learning and co-training were compared against LDS in breast cancer and HCC. LDS Matlab implementation was downloaded from [41]. Tables 1 and 2 show the size of the training and testing sets for each cancer type according to its subtypes. All miRNA and gene expression profiles were downloaded from NCBI [36]. Moreover, table 3 shows sample size and miRNA/gene numbers in the unlabeled sets.

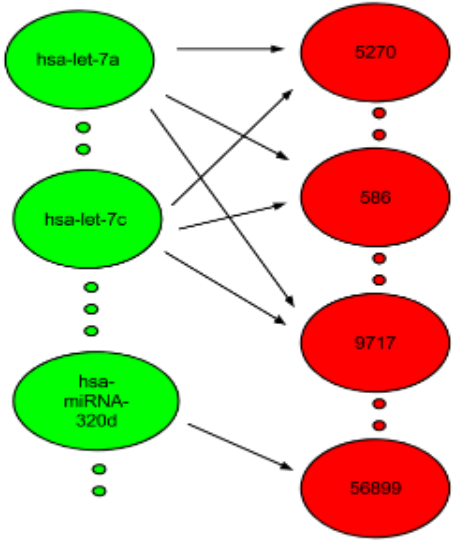

Fig 2. miRNAs and their target genes are related by a many to many relationship. The first column represents miRNAs and the second column represents target genes ids.

---

***Co-training*** ( $L_{miRNA}$, $U_{miRNA}$, $L_{gene}$, $U_{gene}$, α)

**Inputs**: miRNA expression profile of labeled set ($L_{miRNA}$), miRNA expression profile of unlabeled sets ($U_{miRNA}$), gene expression profile of labeled set ($L_{gene}$), gene expression profile of unlabeled sets ($U_{gene}$) and confident threshold (α)

**Outputs**: Two classifiers ($C_{miRNA}$) and ($C_{gene}$) that can separately classify cancer samples.

**Begin**

  *Repeat* {
  1. $C_{miRNA}$ = TrainClassifier ($L_{miRNA}$)
  2. $C_{gene}$ = TrainClassifier ($L_{gene}$)
  3. $L'_{miRNA}$ = Classify ($C_{miRNA}$, $U_{miRNA}$)
  4. $Lα_{miRNA}$ = ChooseMostConfident($L'_{miRNA}$, α)
  5. $L'_{gene}$ = Classify ($C_{gene}$, $U_{gene}$).
  6. $Lα_{gene}$ = ChooseMostConfident($L'_{gene}$, α).
  7. $L_{miRNA}$ = $L_{miRNA}$ U ConverToMiRNAs($Lα_{gene}$)
  8. $L_{gene}$ = $L_{gene}$ U ConverToGenes($Lα_{miRNA}$)
  } **Until** (no improvement in classification accuracy or reaching max iterations)

**End**

Fig 3. Pseudo code of co-training approach.

---

Table 1. Training and testing samples size for breast cancer and HCC subtypes using miRNA expression. (NM = non-metastatic and M = metastatic)

| Type | Breast Cancer (GSE15885) | | | | HCC (GSE6857) | |
|---|---|---|---|---|---|---|
| Subtype | ER+/Her2- | ER-/Her2+ | ER-/Her2- | ER+/Her2+ | NM | M |
| Train | 8 | 2 | 5 | 1 | 193 | 62 |
| Test | 7 | 2 | 4 | 0 | 162 | 65 |

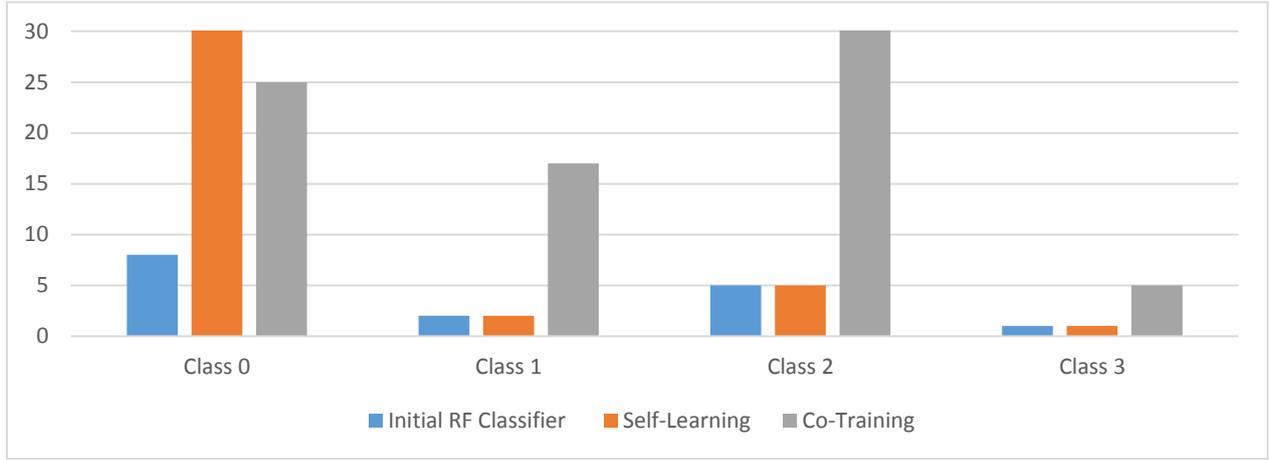

Fig 4. Training data size comparison of initial RF classifier against self-learning and co-training on breast cancer using miRNA expression sets. (Class 0 is ER+/Her2-, class 1 is ER-/Her2+, class 2 is ER-/Her2- and class3 is ER+/Her2+)

Table 2. Training and testing sample size for breast cancer, HCC and lung cancer subtypes using gene expression. (NM = non-metastatic, M = metastatic, A = adenocarcinoma and S = squamous)

| Type | Breast Cancer (GSE20713) | | | | HCC | | Lung | |
|---|---|---|---|---|---|---|---|---|
| Sub Type | ER+/Her2- | ER-/Her2+ | ER-/Her2- | ER+/Her2+ | NM | M | A | S |
| Train | 17 | 9 | 14 | 4 | 98 | 118 | 87 | 41 |
| Test | 17 | 8 | 14 | 4 | 98 | 120 | 86 | 40 |

Table 3. Sample size and miRNA/gene numbers of unlabeled sets.

| | Sample Size | miRNA/gene Numbers |
|---|---|---|
| Breast Cancer (GSE26659) | 94 | 237 miRNAs |
| Breast Cancer (GSE35412) | 35 | 378 miRNAs |
| Breast Cancer (GSE16179) | 19 | 54675 genes |
| Breast Cancer (GSE22865) | 12 | 54675 genes |
| Breast Cancer (GSE32161) | 6 | 54675 genes |
| HCC (GSE10694) | 78 | 121 miRNAs |
| HCC (GSE15765) | 90 | 42718 genes |
| Lung Cancer (GSE42127) | 176 | 48803 genes |

Table 4. Precision, recall and F1-measure for breast cancer subtypes RF classifiers using miRNA expression and gene expression dataset.

| | miRNAs based classifier | | | Genes based classifier | | |
|---|---|---|---|---|---|---|
| | P | R | F1 | P | R | F1 |
| Baseline classifier | 28.9 | 53.9 | 37.7 | 40.8 | 44.2 | 42.5 |
| Self-learning iteration 1 | 31.4 | 53.9 | 39.7 | **54.6** | **58.1** | **56.3** |
| Self-learning iteration 2 | 46.8 | 61.5 | 53.2 | - | - | - |
| Co-training | **56.9** | **61.5** | **59.1** | 44.7 | 53.5 | 48.7 |

Table 5. Precision, recall and F1-measure for breast cancer subtypes SVM classifiers using miRNA expression and gene expression dataset.

| | miRNAs based classifier | | |
|---|---|---|---|
| | P | R | F1 |
| Baseline SVM classifier | 24.5 | 38.5 | 29.9 |
| LDS | 42.3 | 30.8 | 35.6 |
| Self-Learning | 31.4 | 53.8 | 39.7 |
| Co-Training | **62.2** | **61.5** | **61.9** |

A. *Breast Cancer*

Breast cancer is a heterogeneous disease that has a range of phenotypically distinct tumor types. This heterogeneity has an underlying spectrum of molecular alterations and initiating events that was found clinically through a diversity of disease presentations and outcomes [21]. Due to the complexity of this type of tumor, there is a demand for an efficient classification of breast cancer samples.

For breast cancer, both self-learning and co-training are used. Self-learning was applied for both miRNA and gene based classifiers. For sample classification using miRNA expression dataset, an initial breast cancer subtype labeled dataset (GSE15885) was used to build an initial cancer subtype classifier. The initial classifier was then used to predict the labels of the unlabeled breast cancer subtypes (GSE26659 and GSE35412).Two iterations were performed over the two unlabeled datasets. The confident samples, the ones with classification confidence ($\alpha$) greater than 0.9 were added to the training dataset and the subtype classifier was re-trained. The same operation was repeated for sample classification using gene expression dataset where the initial dataset (GSE20713) was used to build an initial classifier and the unknown subtype breast cancer (GSE16179) was used to enrich it. Table 4 shows the precision, recall and F1-measure enhancement against the RF classifier. The results show 12% improvement in F1-measure of breast cancer subtype classifier using miRNA expression profiles and 6% improvement in F1-measure of breast cancer subtype classifier using gene expression profiles. Moreover, table 5 shows the enhancement over SVM and LDS classifiers, only miRNA expression profiles were used in this comparison as LDS requires a lot of memory and thus was unable to use with large number of genes. The table shows that self-learning achieved 10% improvement in F1-measure over SVM classifier and 4% improvement in F1-measure over LDS classifier.

Co-training was evaluated in breast cancer subtypes in both miRNA expression and gene expression. To enhance sample classification using miRNA expression, one labeled miRNA expression dataset (GSE15885) is used. One labeled gene expression dataset (GSE20713) and three unlabeled gene expression datasets (GSE16179, GSE22865 and GSE32161) are mapped into miRNA expression values (as explained in subsection B of section III). In addition, to enhance sample classification using gene expression, one labeled gene expression dataset (GSE20713) is used. One labeled miRNA expression dataset (GSE15885) and two unlabeled miRNA expression datasets (GSE26659 and GSE35412) are mapped into gene expression values and added to the gene training dataset. Table 4 shows the significant improvements in F1-measure using co- training over RF classifier. Increments up to 21% and 8% in F1-measure are observed when using miRNA expression profiles and gene expression profiles respectively. Moreover, table 5 shows the enhancement of co-training over SVM and LDS classifiers, co-training was able to enhance the F1-measure by around 25% over the LDS classifier.

To have a closer look on the behavior of the methods, the number of training data at each class is determined and shown at figure 4. The figure shows that co-training was able to enrich the training data in all 4 classes which is reflected in the highest improvement in the results and self- learning was able to enrich that training set in class 0.

B. HCC

Hepatocellular carcinoma (HCC) represents an extremely poor prognostic cancer that remains one of the most common and aggressive human malignancies worldwide ([37], [38]). Metastasis is a complex process that involves multiple alterations ([39], [40]), that is why discriminating metastasis and non-metastasis HCC is a challenging problem.

For HCC, both self-learning and co-training approaches were evaluated to discriminate between metastatic and non-metastatic HCC. The self-learning steps are applied using GSE6857 as an initial labeled miRNA expression dataset and GSE10694 as the unlabeled subtypes HCC samples. Also, GSE36376 was used as initial labeled gene expression datasets and GSE15765 as the unlabeled subtypes HCC samples. For co-training, to enhance sample subtype classifier using miRNA expression, one labeled miRNA expression dataset (GSE6857) is used. One labeled gene expression dataset (GSE36376) and one unlabeled gene expression datasets (GSE15765) are mapped into miRNA expression values and added to the miRNA training datasets and the sample subtype classifiers are re-trained. Also, in order to enhance the sample classification using gene expression, one labeled gene expression dataset (GSE36376) is used. One labeled miRNA expression dataset (GSE6857) and one unlabeled miRNA expression datasets (GSE10694) are mapped into gene expression datasets and added to the gene training dataset.

Table 6 shows detailed results for HCC subtype classification using RF core classifier, there is around 10% improvement in precision of HCC metastasis class using miRNA expression sets and around 2% in F1-measure using gene expression sets. Moreover, table 7 shows the improvement of the techniques over SVM and LDS classifiers. Co-training achieved 5% enhancement in recall over SVM classifier and 6% enhancement in F1-measure over LDS classifier. The improvement in HCC is less than breast cancer as in breast cancer the number of used unlabeled sets are larger. Also, the overlapping between the miRNAs and genes between the initial set and the added sets is an important factor. In order to understand why enhancements in breast cancer were more significant, the number of overlapping miRNAs and genes is calculated. Tables 8 and 9 show that the higher the overlap between the miRNAs and genes of the initial set and those of the added sets, the higher the improvements become.

C. Lung Cancer

Lung cancer is the leading cause of cancer-related death in both men and women worldwide, it results in over 160,000 deaths annually [8]. Only self-learning using gene expression dataset was evaluated in lung cancer as no labeled miRNA expression dataset was found on the web. The aim of the cancer subtype classifier is to discriminate between adenocarcinoma and squamous lung cancer subtypes. The labeled gene expression dataset (GSE41271) was used to build an initial classifier and the unlabeled gene expression dataset (GSE42127) was used to enhance it. Table 10 shows the enhancement achieved by self-learning, which is around

3% improvement in F1-measure of squamous lung cancer class.

Table 6. Results of HCC RF subtype classifiers using miRNA/gene expression dataset.

|  | Class NM | | | Class M | | | Weighted Evaluation | | |
|---|---|---|---|---|---|---|---|---|---|
|  | P | R | F1 | P | R | F1 | P | R | F1 |
| miRNA initial classifier | 75 | 93 | 83 | 59 | 24 | 34 | 70 | 73 | 72 |
| miRNA self-learning | 76 | **95** | 84 | **69** | 24 | 36 | 74 | 75 | 74 |
| miRNA co-training | **76.6** | 95 | **84.9** | 69 | **27** | **39** | 74 | 75 | 75 |
| Genes initial classifier | 95 | **98** | 96 | **98** | 95 | 97 | 96 | 96 | 96 |
| Genes self-learning | **100** | 96 | **98** | 97 | **100** | **98** | **98** | **98** | **98** |

Table 7. Results of HCC subtype SVM classifiers using miRNA expression dataset.

|  | P | R | F1 |
|---|---|---|---|
| Baseline SVM classifier | 66.5 | 66.1 | 66.3 |
| LDS | 61.2 | 66.9 | 63.9 |
| Self-Learning | 62.2 | 61.5 | 61.9 |
| Co-Training | **67.7** | **71.4** | **69.5** |

Table 8. The number of overlapping miRNAs and genes between initial datasets and added datasets in breast cancer.

|  | miRNAs initial dataset (GSE15885) | Genes initial dataset (GSE20713) |
|---|---|---|
| GSE15885 | 336 | 7 |
| GSE26659 | 124 | 7 |
| GSE35412 | 183 | 7 |
| GSE20713 | 157 | 54676 |
| GSE16179 | 157 | 54676 |

Table 9. The number of overlapping miRNAs and genes between initial datasets and added datasets in HCC.

|  | miRNAs initial dataset (GSE6857) | Genes initial dataset (GSE36376) |
|---|---|---|
| GSE10694 | 36 | - |
| GSE36376 | 52 | 47323 |
| GSE15765 | 52 | 37282 |

Table 10. Results of lung cancer RF subtypes classifiers using gene expression dataset.

|  | Class A | | | Class S | | | Weighted Evaluation | | |
|---|---|---|---|---|---|---|---|---|---|
|  | P | R | F1 | P | R | F1 | P | R | F1 |
| Genes initial classifier | 83 | 95 | 89 | 83 | 54 | 65 | 83 | 83 | 83 |
| Genes self-learning | **84** | **96** | **90** | **87** | **56** | **68** | **85** | **85** | **85** |

## V. CONCLUSION

In this paper, two semi-supervised machine learning approaches were adapted to classify cancer subtype based on miRNA and gene expression profiles. They both exploit the expression profiles of unlabeled samples to enrich the training data. The miRNA-gene relation is additionally used to enhance the classification in co-training. Both self-learning and co-training approaches improved the accuracy compared to Random Forests and SVM as baseline classifiers. The results show up to 20% improvement in F1-measure in breast cancer, 10% improvement in precision in metastatic HCC cancer and 3% improvement in F1-measure in squamous lung cancer. Co-Training also outperforms Low Density Separation (LDS) approach by around 25% improvement in F1-measure in breast cancer.